\documentstyle[twocolumn]{article}

\makeatletter
\def\@normalsize{\@setsize\normalsize{12pt}\xpt\@xpt
\abovedisplayskip 10pt plus2pt minus5pt\belowdisplayskip \abovedisplayskip
\abovedisplayshortskip \z@ plus3pt\belowdisplayshortskip 6pt plus3pt
minus3pt\let\@listi\@listI} 
\def\subsize{\@setsize\subsize{12pt}\xipt\@xipt}
\def\section{\@startsection {section}{1}{\z@}{24pt plus 2pt minus 2pt}
{12pt plus 2pt minus 2pt}{\large\bf}}
\def\subsection{\@startsection {subsection}{2}{\z@}{12pt plus 2pt minus 2pt}
{12pt plus 2pt minus 2pt}{\subsize\bf}}
\makeatother

\newtheorem{Fact}{Fact}
\newtheorem{Facts}{Facts}
\newtheorem{Proposition}{Proposition}

\newtheorem{Theorem}{Theorem}

\newcommand{\type}{ \,{\bf :}\, }

\newcommand{\memberof}{\,{\in}\,}
\newcommand{\define}{\stackrel{{\rm df}}{=}}

\newcommand{\interior}[2]{{#1}^{\circ}_{#2}}
\newcommand{\closure} [2]{{#1}^{\bullet}_{#2}}

\newcommand{\product}[2]{ {#1} {\times} {#2} }

\newcommand{\pair}[2]{\langle #1,#2 \rangle}

\newcommand{\triple}[3]{\mbox{$ \langle #1,#2,#3 \rangle $}}

\newcommand{\quintuple}[5]{\mbox{$ \langle #1,#2,#3,#4,#5 \rangle $}}

\newcommand{\derivedir}[2]{ {#1}^{\Rightarrow}_{#2} }
\newcommand{\deriveinv}[2]{ {#1}^{\Leftarrow}_{#2} }

\title{ \Large{\bf Formal Concept Analysis with Many-sorted Attributes}\normalsize }

\author{
\begin{tabular}[t]{c@{\hspace{25pt}}c}
        Robert E.\ Kent
         & John Brady \\ \\
        Dept.\ of Computer \& Information Science
         & \hspace{13pt}Systematics Information Services, Inc.\hspace{13pt} \\
        University of Arkansas at Little Rock
         &  Little Rock, Arkansas, 72212 \\
        (rekent@logos.ualr.edu)
         & \\ \\
        \multicolumn{2}{c}{{KEYWORDS:} formal context, concept lattice, network} \\
        \multicolumn{2}{c}{of constraints, distributed relation, satisfaction}
\end{tabular}}

\date{}

\begin{document}
\maketitle
\thispagestyle{empty}

\subsection*{\centering Abstract}
{\em
This paper unites two problem-solving traditions in computer science:
(1) constraint-based reasoning; and
(2) formal concept analysis.
For basic definitions and properties of networks of constraints,
we follow the foundational approach of Montanari and Rossi \cite{MR88}.
This paper advocates distributed relations
as a more semantic version of networks of constraints.
The theory developed here uses the theory of formal concept analysis,
pioneered by Rudolf Wille and his colleagues \cite{W92},
as a key for unlocking the hidden semantic structure within distributed relations.
Conversely,
this paper offers distributed relations
as a seamless many-sorted extension to
the formal contexts of formal concept analysis.
Some of the intuitions underlying our approach
were discussed in a preliminary fashion 
by Freuder and Wallace 
\cite{FW92}.
}


\section*{Introduction}

The fundamental model-theoretic structures in this paper are distributed relations.
Distributed relations are identified with
the constraint systems of object-oriented constraint languages,
which play the roles of code and data abstractions,
and subsume classes, instance variables and methods \cite{H92}.
Object state is defined by solving constraints.
Constraints can specify various consistency requirements of objects.
External hierarchical structure is definable by summation of constraints.
Internal hierarchical structure is definable via formal concept analysis
in terms of the concept lattice of associated formal satisfaction contexts.
Formal contexts are shown to be equivalent to single-sorted distributed relations.
The formal satisfaction context of any distributed relation
is the inverse image of the relation along the projection domain morphism.
Concept lattices form a hierarchical clustering of objects.
This hierarchy represents all implications between constraints
(an intensional logic).

The knowledge representation research group here at the University of Arkansas
is currently using formal concept analysis in the area of natural language modelling.
More specifically,
we are using the conceptual hierarchies of concept lattices to investigate and represent
whole language semantic space as incorporated in dictionaries and thesauri.


\section{Formal Concept Analysis}

Formal concept analysis is a new approach to formal logic and knowledge representation
initiated by Rudolf Wille \cite{W82,W92}.
Formal concept analysis starts with the primitive notion of a formal context.
A {\em (order-theoretic) formal context\/} is a triple
${\cal C} = \triple{G}{M}{I}$
consisting of two posets $G = \pair{G}{\leq_G}$ and $M = \pair{M}{\leq_M}$
and a binary relation $I \subseteq \product{G}{M}$ between $G$ and $M$
which respects order:
$g_1 \leq_G g_2$, $g_2{I}m$ imply $g_1{I}m$; and
$g{I}m_1$, $m_1 \leq_M m_2$ imply $g{I}m_2$.
Intuitively, the elements of $G$ are thought of as {\em entities\/} or {\em objects\/},
the elements of $M$ are thought of as {\em properties\/}, {\em characteristics\/} or {\em attributes\/}
that the entities might have,
and $g{I}m$ asserts that ``object $g$ {\em has\/} attribute $m$.''
This definition extends the original notion of formal context
which was given in a set-theoretic realm.
Theoretically,
there are strong reasons for enriching and extending
to an order-theoretic framework in an order-theoretic setting.
Practically,
this enrichment offers the advantage of greater expressibility
for both the system analyst and system designer,
since it allows one to either describe or specify order constraints
on both objects and attributes.

The collection
${\rm Cxt}_G^M$ of all contexts
with a fixed object set $G$ and a fixed attribute set $M$
is a poset with the subset inclusion pointwise order $I \subseteq I'$.
As attribute sets $M$ are allowed to vary,
we collect together all contexts with fixed object set $G$
in the poset
${\bf Cxt}_G \define \coprod_M {\rm Cxt}_G^M$.
We can define direct and inverse image operators on contexts 
along any function $\phi \type G_2 \rightarrow G_1$ between object sets
as follows.
\begin{description}
	\item[{[Direct image]}] Let $\triple{G_2}{I_2}{M}$ be any context.
		Define its direct image along $\phi$ to be the context
		$\triple{G_1}{\phi^{\rm op} \circ I_2}{M}$.
		Direct image is a monotonic function
		$\phi_\ast \type {\bf Cxt}_{G_2} \rightarrow {\bf Cxt}_{G_1}$.
	\item[{[Inverse image]}] Let $\triple{G_1}{I_1}{M}$ be any context.
		Define its inverse image along $\phi$ to be the context
		$\triple{G_2}{\phi \circ I_1}{M}$.
		Inverse image is a monotonic function
		$\phi^\ast \type {\bf Cxt}_{G_1} \rightarrow {\bf Cxt}_{G_2}$.
\end{description}
Contextual flow is adjoint:
for any function between object sets
$\phi \type G_1 \rightarrow G_2$,
direct and inverse image form an adjoint pair of monotonic functions
\[ \phi_\ast \;\dashv\; \phi^\ast . \]

Formal concept analysis is based upon the understanding that
a concept is a unit of thought consisting of two parts:
its extension and its intension
(Woods \cite{W91} also advocates conceptual intensions).
Within the restricted scope of a formal context,
the {\em extent\/} of a concept is a subset $\phi \subseteq G$ consisting of all objects belonging to the concept,
whereas
the {\em intent\/} of a concept is a subset $\psi \subseteq M$ which includes all attributes shared by the objects.
A concept of a given context will consist of an extent/intent pair
\[ ({\phi},{\psi}). \]

Of central importance in concept construction are two {\em derivation operators\/}
which define the notion of ``sharing'' or ``commonality''.
For any subset of objects $\phi \subseteq G$
the direct derivation along $I$ is
$\derivedir{\phi}{I} \define \{ m \in M \mid g{I}m \mbox{ for all } g \in \phi \}$.
For any subset of attributes $\psi \subseteq M$
the inverse derivation along $I$ is
$\deriveinv{\psi}{I} \define \{ g \in G \mid g{I}m \mbox{ for all } m \in \psi \}$.
These two derivation operators form an adjointness (generalized inverse relationship)
$\derivedir{\phi}{I} \supseteq \psi$ iff $\phi \subseteq \deriveinv{\psi}{I}$.
To demand that a concept $({\phi},{\psi})$ be determined by its extent and its intent
means that this adjointness should be
a strict inverse relationship at the extent/intent pair $({\phi},{\psi})$:
the intent should contain precisely those attributes shared by all objects in the extent
$\derivedir{\phi}{I} = \psi$,
and vice-versa,
the extent should contain precisely those objects sharing all attributes in the intent
$\phi = \deriveinv{\psi}{I}$.
Concept extents are identical to the closed subsets of the closure operator
$\closure{\phi}{I} \define \deriveinv{(\derivedir{\phi}{I})}{I}$
defined by the adjointness of derivation,
and
concept intents are identical to the open subsets of the interior operator
$\interior{\psi}{I} \define \derivedir{(\deriveinv{\psi}{I})}{I}$
defined by this adjointness.

The collection of all concepts is ordered by generalization-specialization.
One concept is more specialized (and less general) than another
$(\phi,\psi) \leq_{\rm L} (\phi',\psi')$
when its intent contains the other's intent
$\psi \supseteq \psi'$,
or equivalently,
when the opposite ordering on extents occurs
$\phi \subseteq \phi'$.
Concepts with this generalization-specialization ordering
form a concept hierarchy for the context.
The concept hierarchy is a complete lattice ${\bf L}({\cal C})$
called the {\em concept lattice\/} of ${\cal C}$.

The meets and joins of concepts in ${\bf L}({\cal C})$ can be described
in terms of unions, intersections, intent interior, and extent closure as follows:
\footnotesize
\begin{center}
	$\begin{array}{r@{\;=\;}l}
		\bigwedge_{k \in K} (\phi_k,\psi_k)
		& \left( \bigcap_{k \in K} \phi_k, \closure{(\bigcup_{k \in K} \psi_k)}{I} \right)
		\\
		\bigvee_{k \in K} (\phi_k,\psi_k)
		& \left( \closure{(\bigcup_{k \in K} \phi_k)}{I}, \bigcap_{k \in K} \psi_k \right) .
	 \end{array}$
\end{center}
\normalsize
The join of a collection of concepts
represents the common properties (shared characteristics)
of the concepts,
and the top of the concept hierarchy represents the universal concept
whose extent consists of all objects.
When extended to distributed relations,
the meet of a collection of concepts
corresponds to the natural join from relational theory,
and the bottom of the concept hierarchy represents all solutions that satisfy the constraints
(the solution-set concept).

In this paper we give arguments that Wille's original notion of formal context,
although quite appealing in its simplicity,
now should be extended to distributed relations.
Such a generalization and abstraction of formal contexts offers a powerful approach
for the representation of knowledge and the reasoning about constraints.


\section{Distributed Relations}

The basic parameter in relational theory is a sorted domain.
Let $N$ be a set of sorts.
An $N$-sorted domain $D$ is an $N$-indexed collection of posets
$D = \{ D_a \mid a \memberof N \}$.
A single-sorted domain ($N = 1$) is just a poset.
Let $\wp N$ denote the set of all finite subsets of $N$.
A finite subset of sorts $U \memberof \wp N$ is called
either an {\em arity} or an {\em elementary relational scheme}.
Given any scheme $U = \{ a_1,...,a_n \}$,
the $U$-th power of $D$ is defined by
$D^U = D_{a_1} {\times} \ldots {\times} D_{a_n}$.
The empty power is $D^\emptyset = {\rm 1}$, a canonical singleton poset.
Any sort subset inclusion $U \subseteq V$
defines an obvious projection monotonic function
$D^V \stackrel{\pi_{VU}}{\longrightarrow} D^U$.

The basic variable quantity in relational theory is a relation.
An $N$-sorted {\em relation\/} over domain $D$ with scheme $U \memberof \wp N$,
an $\pair{N}{D}$-{\em relation\/},
is a closed-below subset $R \subseteq D^U$.
We use the declaration notation $R \type U$ to denote this.
For any two $\pair{N}{D}$-relations $S \type V$ and $R \type U$,
a {\em projective containment\/} condition exists from $S \type V$ to $R \type U$,
written $S \type V \leq R \type U$,
when
(1) $U \subseteq V$, and
(2) $\pi_{VU}(S) \subseteq R$
    or equivalently
    $S \subseteq \pi_{VU}^{-1}(R)$.
Containment conditions
(either these simple projective containment conditions
or nontrivial dependency containment conditions)
are examples of {\em internal constraints\/} on data.
Internal constraints are a particular specification form for semantics.

Let ${\rm Rel}_{N,D}$ denote the collection of all $\pair{N}{D}$-relations.
This forms a poset with projective containment order.
For a single-sorted domain $D$,
${\rm Rel}_{{\rm 1},D}$ is the powerset of $D$ plus the two relations
$\emptyset \type \emptyset$ and ${\rm 1} \type \emptyset$.
The {\em underlying scheme function\/}
$p_{N,D} \type {\rm Rel}_{N,D} \longrightarrow {\wp N}^{\rm op}
         \type (R \type U) \mapsto U$
is monotonic by the definition of projective containment.
\begin{Fact}
	${\rm Rel}_{N,D}$ is a complete lattice:
	infimums are natural joins,
	the top is ${\rm 1} \type \emptyset$, and
	the bottom is $\emptyset \type N$.
\end{Fact}

Given a set of sorts $N$,
a {\em distributed relational scheme\/} over $N$
is a pair $\pair{E}{\tau}$ consisting of
(1) a preorder of {\em relation\/} ({\em predicate\/}) {\em symbols or indices\/}
$E = \pair{E}{\leq_E}$, and
(2) a monotonic function
$\tau \type E \rightarrow {\wp N}^{\rm op}$
assigning elementary relational schemes to relation indices.
The triple $\Omega = \triple{E}{\tau}{N}$
is called a {\em first order signature\/} (without multiplicities).
The signature is discrete
when
the preorder of relation symbols is the identity preorder $E = \pair{E}{=_E}$.
In this case a signature is the same as a hypergraph.
The signature is single-sorted
when
the set of sorts is a singleton set $N = {\rm 1}$.
In this case the signature
$\Omega = \triple{E}{{\bf !}}{{\rm 1}}$
is equivalent to a preorder $E$.

Constraint-based reasoning
is based upon the primitive notion of a distributed relation.
Given a fixed signature $\Omega = \triple{E}{\tau}{N}$,
an $N$-{\em sorted distributed relation\/} with scheme $\pair{E}{\tau}$,
an $\Omega$-{\em relation\/},
is a pair $\pair{D}{R}$ consisting of
(1) an $N$-sorted domain $D = \{ D_a \mid a \memberof N \}$,
and
(2) a monotonic function $R \type E \longrightarrow {\rm Rel}_{N,D}$
assigning relations to relation symbols and
projective containment conditions to $E$-inequalities,
where $R$ is compatible with the scheme $\pair{E}{\tau}$
in the sense that
$R \cdot p_{N,D} = \tau$.
The latter equality defines when $R$ has distributed scheme $\tau$.
\begin{Facts} \mbox{ }
\begin{enumerate}
	\item Order-theoretic formal contexts
		${\cal C} = \triple{G}{M}{I}$
		are (equivalent to)
		distributed relations for single-sorted signatures.
	\item Networks of constraints \cite{MR88}
		${\cal N} = \quintuple{N}{\tau}{E}{D}{R}$
		are defined to be
		distributed relations with discrete signatures.
\end{enumerate}
\end{Facts}
For formal contexts as distributed relations,
(1) the signature is the preorder of attributes
    $\Omega = \triple{M}{{\bf !}}{{\rm 1}}$,
(2) the single domain is the set of objects $D = G$, and
(3) the relation assignment
    $R \type M \longrightarrow {\rm Rel}_{{\rm 1},G}
    = \wp G + \{ \emptyset \type \emptyset, {\rm 1} \type \emptyset \}$
    assigns 
    corresponding columns of the incidence relation to attributes
    $R \type a \mapsto Ia \subseteq G$,
    so that each attribute represents a unary constraint.
A network of constraints is a combinatorial construct,
{\em not\/} a semantic algebraic construct.
Distributed relations
provide a semantic extension of networks of constraints
by specifying projective containment conditions between constraints.
Projective containment conditions help optimize the size of distributed relations.

For a fixed sorted domain $\pair{N}{D}$
the triple $R \type \pair{E}{\tau}$ is called a
{\em distributed\/} $\pair{N}{D}$-{\em relation\/}.
For a fixed {\em signed domain\/} $\pair{\Omega}{D}$,
where $\Omega = \triple{E}{\tau}{N}$ is a signature
and $D$ is an $N$-sorted domain,
that is,
for a fixed sorted domain $\pair{N}{D}$ and any relational scheme $\pair{E}{\tau}$,
the collection
${\rm Rel}_{N,D}^{E,\tau}$ of all distributed $\pair{N}{D}$-relations with scheme $\pair{E}{\tau}$
is a poset
(actually, a complete Boolean algebra) with the pointwise order
$R \leq S$ when $R_e \subseteq S_e$
for all relation indices $e \memberof E$.
If $E = {\rm 1}$ a singleton,
then $\tau \type {\rm 1} \longrightarrow {\wp N}^{\rm op}$
is essentially an elementary relational scheme $U \memberof {\wp N}^{\rm op}$,
and ${\rm Rel}_{N,D}^{E,\tau} = \wp {D^U}$
the complete Boolean algebra of (nondistributed) $\pair{N}{D}$-relations
with scheme $U$.
As the scheme $\pair{E}{\tau}$ varies,
the posets ${\rm Rel}_{N,D}^{E,\tau}$ form our most basic semantic domains,
and have syntactically specified join and cojoin transformations
${\rm Rel}_{N,D}^{E,\tau} \longrightarrow {\rm Rel}_{N,D}^{F,\sigma}$
defined between them.
As distributed schemes $\pair{E}{\tau}$ are allowed to vary,
we collect together all distributed $\pair{N}{D}$-relations
in the poset
${\bf Rel}_{N,D} \define \coprod_{E,\tau} {\rm Rel}_{N,D}^{E,\tau}$.


\section{Domain Morphisms}

The process of varying (weakening or strengthening)
constraint satisfaction problems \cite{FW92}
involves two senses  (external/internal)
and      two options (syntactic/semantic) within each sense.
\begin{enumerate}
	\item External sense: vary sorted domains.
	\begin{enumerate}
		\item Remove variables (domain indices): changing $N$ to $N'$,
			where $N \supseteq N'$.
		\item Enlarge variable domains:
			changing $\{ D_i \mid i \memberof N \}$ to $\{ D'_i \mid i \memberof N \}$,
			where $D_i \subseteq D'_i$ for all $i \memberof N$.
	\end{enumerate}
	\item Internal sense: vary distributed relations.
	\begin{enumerate}
		\item Remove constraints (relation indices): changing $E$ to $E'$,
			where $E \supseteq E'$.
		\item Enlarge constraint relations:
			changing $\{ R_e \mid e \memberof E \}$ to $\{ R'_e \mid e \memberof E \}$,
			where $R_e \subseteq R'_e$ for all $e \memberof E$.
			As noted by \cite{FW92}
			when $R'_i = \top = D^I$ the top unconstraining relation,
			the constraint at $e$ has effectively been removed;
			so that syntactic variability can be effected to some extent by semantic variability.
	\end{enumerate}
\end{enumerate}
The external sense for varying constraint specifications
is formally defined as direct/inverse image along domain morphisms.
A {\em morphism of sorted domains\/}
$\pair{f}{\phi} \type \pair{N_1}{D_1} \rightarrow \pair{N_2}{D_2}$
is a pair consisting of
\begin{enumerate}
	\item a function
	    $f \type N_1 \rightarrow N_2$
	    between sort sets, and
	\item a $\wp N_1$-indexed collection
	    $\phi = \{ D_2^{fU} \stackrel{\phi_U}{\rightarrow} D_1^U \mid U \subseteq N_1 \}$
	    of monotonic functions between domains,
	    which satisfy
	    $\phi_V \cdot \pi_{VU} = \pi_{{fV}{fU}} \cdot \phi_U$
	    for every pair $V \supseteq U$ of subsets of $N_1$.
\end{enumerate}
Domain morphisms specify adjoint (direct/inverse image) relational flow.
\begin{description}
	\item[{[Direct image]}] Let $R_2 \type \pair{E_2}{\tau_2}$ be any distributed
		$\pair{N_2}{D_2}$-relation.
		Define distributed scheme $\pair{E_1}{\tau_1}$ to be the pullback
		of scheme $\pair{E_1}{\tau_1}$ along the direct image sort map
		$\wp f \type \wp N_1 \rightarrow \wp N_2$
		with fibered product
		$E_1 \define \{ (e_2,U) \mid e_2 \memberof E_2, U {\subseteq} N_1, \tau_{2,e_2} {=} fU \}$
		and projection function
		$\tau_1 \type (e_2,U) \mapsto U$.
		Define distributed relation
		$R_{1,(e_2,U)} \define \phi_U(R_{2,e_2})$
		for any $(U,e_2) \in E_1$
		to be the direct image along domain function
		$D_2^{\tau_{2,e_2}} = D_2^{fU} \stackrel{\phi_U}{\longrightarrow} D_1^U$.
		Finally,
		define the direct image distributed $\pair{N_1}{D_1}$-relation
		${\pair{f}{\phi}}_\ast(R_2 \type \pair{E_2}{\tau_2})
		  \define R_1 \type \pair{E_1}{\tau_1}$.
		Direct image is a monotonic function:
		\[ {\pair{f}{\phi}}_\ast \type {\bf Rel}_{N_2,D_2} \rightarrow {\bf Rel}_{N_1,D_1}. \]
	\item[{[Inverse image]}] Let $R_1 \type \pair{E_1}{\tau_1}$ be any distributed
		$\pair{N_1}{D_1}$-relation.
		Define distributed scheme
		$\pair{E_2}{\tau_2}
		 \define \pair{E_1}{\tau_1 {\cdot} {\wp f}^{\rm op}}$.
		Define distributed relation
		$R_{2,e_1} \define \phi_{\tau_{1,e_1}}^{-1}(R_{1,e_1})$
		for any $e_1 \in E_1$
		to be the inverse image along domain function
		$D_2^{\tau_{2,e_1}} = D_2^{f\tau_{1,e_1}} \stackrel{\phi_{\tau_{1,e_1}}}{\longrightarrow} D_1^{\tau_{1,e_1}}$.
		Finally,
		define the inverse image distributed $\pair{N_2}{D_2}$-relation
		${\pair{f}{\phi}}^\ast(R_1 \type \pair{E_1}{\tau_1}) \define R_2 \type \pair{E_2}{\tau_2}$.
		Inverse image is also monotonic:
		\[ {\pair{f}{\phi}}^\ast \type {\bf Rel}_{N_1,D_1} \rightarrow {\bf Rel}_{N_2,D_2}. \]
\end{description}

\begin{Proposition}
	Relational flow is adjoint:
	for any morphism of sorted domains
	$\pair{f}{\phi} \type \pair{N_1}{D_1} \rightarrow \pair{N_2}{D_2}$,
	direct and inverse image form an adjoint pair of monotonic functions
	\[ {\pair{f}{\phi}}_\ast \;\dashv\; {\pair{f}{\phi}}^\ast \]
\end{Proposition}


\section{The Satisfaction Context}

Wille's formal contexts explicitly model
{\em has\/} relationships between objects and attributes (unary constraints).
However,
the more general {\em satisfaction\/} relationships
between object tuples and constraints that we define in this section
associate an order-theoretic formal context
with each distributed relation ${\cal R} = \quintuple{N}{\tau}{E}{D}{R}$
in a very natural fashion.

The basic constituents in constraint-based reasoning
are
tuples of values representing semantic configurations of real-world objects,
and
ways of describing these in terms of constraining relations.
The relationship between tuples and constraints is a derived relationship
called {\em satisfaction\/}.
Intuitively, for any subset $U \subseteq N$
the tuple values in $D^U$,
thought of as the possible state or configuration of entities or objects,
will be called {\em object tuples\/}.
An object tuple may represent the participation of several objects or entities
in a semantic whole.
A tuple $x \memberof D^U$ is said to have {\em tag\/} or {\em arity\/} $U$
and is denoted by $x \type U$.
The set of all nonempty tuples of a distributed relation
can be defined as the disjoint union
$D^{[N]}
 = \coprod_{U \subseteq N} D^U
 \define \bigcup_{U \subseteq N} \product{\{U\}}{(D^U)}$.
This is a partially ordered set with tuples in $D^{[N]}$ ordered by projection:
$y \type V \preceq x \type U$ when $x$ is the projection of $y$;
that is,
when $V \supseteq U$ and $\pi_{VU}(y) = x$.
This tuple projection order is an instance of meronymy,
or whole-part order.
The tuple $x \type U$ is part of the tuple $y \type V$;
The empty tuple $\varepsilon \type \emptyset$
is the top, smallest part in this order.
A full tuple $x \type N$,
whose arity consists of all indices $N$,
is a minimal element in this identification order,
and represents a whole.

The elements of $E$ are thought of as {\em constraints\/}
that the object tuples might satisfy.
A tuple $x \type U$ {\em satisfies\/} a constraint $e$,
denoted by $x \models e$,
when the tuple generalizes to,
or equivalently is a specialization from,
the relation of the constraint;
that is,
when $U \supseteq \tau_e$ and $\pi_{U{\tau_e}}(x) \memberof R_e$.
Satisfaction is a binary relation
${\models} \subseteq \product{D^{[N]}}{E}$
on the set of all tuples,
which respects order:
$y \type V \preceq x \type U$, $x \models e$ imply $y \models e$; and
$x \models e_1$, $e_1 \leq_E e_2$ imply $x \models e_2$.
So satisfaction forms an order-theoretic formal context ${\cal C}_{\cal R}$
called the {\em formal satisfaction context\/} of ${\cal R}$.
This formal context is defined to be the triple
${\cal C}_{\cal R} = \triple{D^{[N]}}{E}{\models}$,
whose contextual objects are tuples of any arity,
whose contextual attributes are relation symbols (constraints) with defined scheme,
and
whose contextual {\em has\/} relationship is the satisfaction relationship.
Zickwolff \cite{Z91} has independently given a similar development for satisfaction.
The attributes in $E$,
being relation symbols with schema,
are many-sorted
--- whence the title of this paper.

In order to define a canonical projection domain morphism
from distributed relations to formal contexts,
we must restrict satisfaction to full scheme tuples
${\models} \subseteq \product{D^N}{E}$.
Let us denote this restricted context by
${\cal C}_{\cal R}^N = \triple{D^N}{E}{\models}$.
This association of the restricted formal satisfaction context
with each distributed relation,
is represented by the relation-to-context passage
\begin{equation}
	{\cal R}\;\;\; \mapsto \;\;\;{\cal C}_{\cal R}^N. \label{passage}
\end{equation}
For any sorted domain $\pair{N}{D}$,
the full $N$th power $D^N$,
the unique constant function $N \stackrel{{\bf !}}{\rightarrow} {\rm 1}$, and
the collection of all projection functions
$\pi = \{ D^N \stackrel{\pi_{N,U}}{\rightarrow} D^U \mid U \subseteq N \}$
form a canonical morphism of sorted domains called {\em projection\/}
\[ \pi_{N,D}
   \define
   \pair{{\bf !}}{\pi} \type \pair{N}{D} \rightarrow \pair{{\rm 1}}{D^N}. \]
\begin{Theorem}
	The satisfaction context of a distributed relation
	is the inverse image along projection
	\[ {\cal C}_{\cal R}^N = \pi_{N,D}^\ast(R \type \pair{E}{\tau}). \]
\end{Theorem}
The satisfaction context of a context
(regarded as a single-sorted distributed relation)
is the original context.
This demonstrates that distributed relations are a proper extension of formal contexts.

Passage~\ref{passage} associates a lattice of semantic concepts
${\bf L}({\cal C}_{\cal R})$ 
with each distributed relation ${\cal R}$,
thus revealing its hidden semantic structure.
The intuitions underlying this semantic structure
can be expressed in terms of partial constraint satisfaction.
The concept lattices of distributed relations
can be directly related to
the problem space interpretation of partial constraint satisfaction in \cite{FW92}.
According to the discussion in \cite{FW92},
a problem space is a set of constraint-satisfaction-problems ordered by their solution sets.
In this paper
we identify the appropriate problem space of a network of constraints,
or more generally a distributed relation ${\cal R}$,
to be the concept lattice ${\bf L}({\cal C}_{\cal R})$
of the formal satisfaction context associated with the relation.
We thus identify constraint-satisfaction-problems of the problem space
with formal concepts in the concept lattice
--- the problem intent is its set of constraints,
and the problem extent is the solution set of the constraints.
We refer here to the lattice elements as problem-concepts.
These problem-concepts represent partial information about objects.
The lattice uses the opposite sense of order than that defined in \cite{FW92}.
Weakening problem constraints or generalizing
means
moving upward in the problem-concept hierarchy.
On the other hand,
moving downward in the problem-concept hierarchy
corresponds to
the monotonic accumulation of partial information about object tuples.

Current and future work involves
(1) the definition of satisfaction algorithms
on the distributed relation's concept lattice,
and
(2) the definition of suitably generalized similarity and distance metrics
on the distributed relation's concept lattice.


\section{Partial Constraint Satisfaction}

The concept lattices of distributed relations
can be directly related to
the problem space interpretation of partial constraint satisfaction in \cite{FW92}.
According to the discussion in \cite{FW92},
a problem space is a set of constraint-satisfaction-problems ordered by their solution sets.
In this paper
we identify the appropriate problem space of a network of constraints,
or more generally a distributed relation ${\cal R}$,
to be
the concept lattice of the formal satisfaction context
associated with the relation ${\bf L}({\cal C}_{\cal R})$.
We thus identify constraint-satisfaction-problems of the problem space
with formal concepts in the concept lattice
--- the problem intent is its set of constraints,
and the problem extent is the solution set of the constraints.
We refer here to the lattice elements as problem-concepts.
These problem-concepts represent partial information about objects.
The lattice uses the opposite sense of order than that defined in \cite{FW92}.
The problem space is restricted by the strong idea of formal concept
--- the constraint set (intent) of each problem
contains {\em all\/} constraints satisfied by
the solution set (extent) of the problem.
These restrictions on allowable problem-concepts make semantic sense
and greatly optimize the solution search process.
The semantic structure of the concept lattice
specifies,
through the strong idea of formal concept,
how constraint-satisfaction-problems should be weakened
and
how allowable problems should be restricted.
Weakening problem constraints or generalizing
means
moving upward in the problem-concept hierarchy.
On the other hand,
moving downward in the problem-concept hierarchy
corresponds to
the monotonic accumulation of partial information about object-tuples.
The top of the hierarchy is unconstrained;
the bottom, whose intent is all constraints,
has the total solution set of the system as its extent.
Maximal constraint satisfaction corresponds to minimality in the lattice hierarchy.
The join of two problem-concepts represents common constraints,
and the meet corresponds to the natural join from relational theory.

Current and future work
involves
the definition of satisfaction algorithms
on the distributed relation's concept lattice
(including generalization of analogs to Freuder and Wallace's satisfaction algorithms
from maximal satisfaction,
which is only one part of the distributed relation's concept lattice,
to the entire lattice),
and
the definition of suitably generalized similarity and distance metrics
on the distributed relation's concept lattice.

\section{Partial Constraint Satisfaction}

The concept lattices of distributed relations
can be directly related to
the problem space interpretation of partial constraint satisfaction in \cite{FW92}.
According to the discussion in \cite{FW92},
a problem space is a set of constraint-satisfaction-problems ordered by their solution sets.
In this paper
we identify the appropriate problem space of a network of constraints,
or more generally a distributed relation ${\cal R}$,
to be
the concept lattice of the formal satisfaction context
associated with the relation ${\bf L}({\cal C}_{\cal R})$.
We thus identify constraint-satisfaction-problems of the problem space
with formal concepts in the concept lattice
--- the problem intent is its set of constraints,
and the problem extent is the solution set of the constraints.
We refer here to the lattice elements as problem-concepts.
These problem-concepts represent partial information about objects.
The lattice uses the opposite sense of order than that defined in \cite{FW92}.
The problem space is restricted by the strong idea of formal concept
--- the constraint set (intent) of each problem
contains {\em all\/} constraints satisfied by
the solution set (extent) of the problem.
These restrictions on allowable problem-concepts make semantic sense
and greatly optimize the solution search process.
The semantic structure of the concept lattice
specifies,
through the strong idea of formal concept,
how constraint-satisfaction-problems should be weakened
and
how allowable problems should be restricted.
Weakening problem constraints or generalizing
means
moving upward in the problem-concept hierarchy.
On the other hand,
moving downward in the problem-concept hierarchy
corresponds to
the monotonic accumulation of partial information about object-tuples.
The top of the hierarchy is unconstrained;
the bottom, whose intent is all constraints,
has the total solution set of the system as its extent.
Maximal constraint satisfaction corresponds to minimality in the lattice hierarchy.
The join of two problem-concepts represents common constraints,
and the meet corresponds to the natural join from relational theory.

Current and future work
involves
the definition of satisfaction algorithms
on the distributed relation's concept lattice
(including generalization of analogs to Freuder and Wallace's satisfaction algorithms
from maximal satisfaction,
which is only one part of the distributed relation's concept lattice,
to the entire lattice),
and
the definition of suitably generalized similarity and distance metrics
on the distributed relation's concept lattice.


\section{Relational Interior}

Any distributed relation contains within it
another distributed relation called its interior
which satisfies all possible projective containment conditions.
Interior is derived from
an interesting adjointness between projection and solution.

On the one hand,
the {\em solution set\/} of a distributed relation
$R \in {\rm Rel}_{N,D}^{E,\tau}$
is defined by
${\Pi}R \define \{ x \memberof D^N \mid x{\models}e \;\mbox{for all}\; e \memberof E \}$.
Two distributed relations $R,S \in {\rm Rel}_{N,D}^{E,\tau}$
are {\em equivalent\/}
when they have the same solution set ${\Pi}R = {\Pi}S$.
Define
${\Pi}^{-1}P \define \{ R \mid {\Pi}R = P \}$
to be the (possibly empty) collection of distributed relations with solution set $P$.
By definition of the solution set ${\Pi}R$,
the projection $\pi_{N{\tau_e}}(x)$ of every tuple $x \memberof {\Pi}R$
is contained in the $e$-th constraint relation $\pi_{N{\tau_e}}(x) \memberof R_e$.
This fact is expressed by the ``counit inequality''
${\pi}({\Pi}R) \subseteq R$.

On the other hand,
let $P \subseteq D^N$ be any subset of full object tuples.
We regard $P$ as a potential solution set.
The {\em projection\/} of $P$
is the $E$-indexed collection of relations
${\pi}P \define \{ \pi_{N{\tau_e}}(P) \mid e \memberof E \}$.
The projection of $P$
clearly satisfies the projective containment conditions
$\pi_{{\tau_e}{\tau_d}}(\pi_{N{\tau_e}}(P)) \subseteq \pi_{N{\tau_d}}(P)$,
and hence defines a distributed relation
$\quintuple{N}{\tau}{E_{\tau}}{D}{{\pi}P}$.
The fact that every tuple in $P$ satisfies every constraint in $E$
is expressed by the ``unit inequality''
$P \subseteq {\Pi}({\pi}P)$.

There are two ways to define a context on full tuples.
\begin{enumerate}
	\item For any subset of full tuples $P \subseteq D^N$,
		define an associated order-theoretic formal context
		${\cal C}_P = \triple{G}{M}{I_P}$
		by
		\footnotesize
		\[ x{I_P}e \mbox{  iff  } x \memberof P. \]
		\normalsize
\end{enumerate}

Since the projection $\pi$ and product $\Pi$ operators are monotonic functions,
we have the following proposition.
\begin{Proposition}
	The projection operator $\pi$
	and
	the solution operator $\Pi$
	form an adjointness or Galois connection
	\[ \pi \dashv \Pi \]
	between distributed relations $R$ and potential solution sets $P$,
	expressed by the logical equivalence:
	${\pi}P \subseteq R$ iff $P \subseteq {\Pi}R$.
\end{Proposition}
By the above adjointness,
projection preserves joins
${\pi}(\cup_k P_k) = \sqcup_k {\pi}(P_k)$,
and solution preserves meets
${\Pi}(\sqcap_k R_k) = \cap_k {\Pi}P_k$.

For any distributed relation
$R \in {\rm Rel}_{N,D}$
with distributed scheme $\pair{E}{\tau}$,
the projection of the solution
$R^{\circ} \define {\Pi}{\pi}(R) = {\pi}({\Pi}(R))$
is called the {\em interior\/} of $R$.
The interior is a distributed relation
which satisfies all possible projective containment conditions.
The above adjointness implies
the existence of minimal equivalent networks of constraints \cite{MR88}.
The interior $R^{\circ}$ is optimal w.r.t.\ satisfaction
--- it is the smallest distributed relation having solution set ${\Pi}R$:
\begin{enumerate}
	\item ${\Pi}(R^{\circ}) = {\Pi}R$; and
	\item if $S$ is an $E$-indexed distributed relations
		having solution set ${\Pi}S = {\Pi}R$,
		then $R^{\circ} \subseteq S$.
\end{enumerate}
In particular,
interior is the minimal distributed relation 
$R^{\circ} = \bigwedge {\Pi}^{-1} {\Pi}R$
equivalent to $R$.
These results were stated and proved in \cite{MR88},
but the proofs were not expressed
by use of the simplifying notion of an adjointness.

\section{Relational Participation}

The previous discussion
leads one to ask why certain tuples are not included in
the interior of a distributed relation $R$.
For any relational constraint index $e \memberof E$
a tuple $x \type \tau_e$
is in the difference
$R_e \setminus R^{\circ}_e$
when it does not participate
in a combining relationship with other tuples
to form a full object tuple that respects all constraints in $E$.
In other words,
it is isolated with respect to $R$.
In this section we will relativize this idea of {\em relational participation\/}
and make it more explicit.

We can emphasize the importance of certain concepts in a concept lattice
${\bf L}({\cal C})$
for a formal context ${\cal C} = \triple{G}{M}{I}$
by specifying this collection of concepts as a suborder
$P \subseteq {\bf L}({\cal C})$.
More precisely,
suppose that
$P \stackrel{\iota}{\rightarrow} {\bf L}({\cal C})$
is an injective monotonic function
which is left adjoint to
${\bf L}({\cal C}) \stackrel{\iota^\dashv}{\rightarrow} P$.
The trivial example of such a suborder is the full lattice,
where $\iota$ and $\iota^\dashv$ are both identity.
The simplest nontrivial example of such a suborder is the principal ideal
${\downarrow}c$ of any concept $c \memberof {\bf L}({\cal C})$,
where $\iota$ is subset inclusion $\iota(c') = c'$ for all $c' \leq_{\rm L} c$
and $\iota^\dashv$ is meet $\iota^\dashv(d) \define c \wedge_{\rm L} d$ for all $d \memberof {\bf L}({\cal C})$.
As a special case, relevant for relational interior, is $c = \bot$.
Define an associated formal context
${\cal C}_P = \triple{G}{M}{I_P}$
by
\footnotesize
\[ g{I_P}m \mbox{  iff  } \exists_{p \in P}\; \gamma(g) \leq_{\rm L} \iota(p) \leq_{\rm L} \mu(m). \]
\normalsize


\section{Domain Morphisms}

In a global sense,
following the discussion in \cite{FW92},
the process of varying (weakening or strengthening) constraint satisfaction problems
involves two senses with 2 options each.
\begin{enumerate}
	\item External sense: vary sorted domains.
	\begin{enumerate}
		\item Remove variables (domain indices): changing $N$ to $N'$,
			where $N \supseteq N'$.
		\item Enlarge variable domains:
			changing $\{ D_i \mid i \memberof N \}$ to $\{ D'_i \mid i \memberof N \}$,
			where $D_i \subseteq D'_i$ for all $i \memberof N$.
	\end{enumerate}
	\item Internal sense: vary distributed relations.
	\begin{enumerate}
		\item Remove constraints (relation indices): changing $E$ to $E'$,
			where $E \supseteq E'$.
		\item Enlarge constraint relations:
			changing $\{ R_e \mid e \memberof E \}$ to $\{ R'_e \mid e \memberof E \}$,
			where $R_e \subseteq R'_e$ for all $e \memberof E$.
			As noted by \cite{FW92}
			when $R'_i = \top = D^I$ the top unconstraining relation,
			the constraint at $e$ has effectively been removed;
			so that syntactic variability can be effected to some extent by semantic variability.
	\end{enumerate}
\end{enumerate}

The external sense for varying constraint specifications
is formally defined as direct/inverse image along domain morphisms.
A {\em morphism of sorted domains\/}
$\pair{f}{\phi} \type \pair{N_1}{D_1} \rightarrow \pair{N_2}{D_2}$
is a pair consisting of
\begin{enumerate}
	\item a function
	    $f \type N_1 \rightarrow N_2$
	    between sort sets, and
	\item a $\wp N_1$-indexed collection
	    $\phi = \{ E^{fU} \stackrel{\phi_U}{\rightarrow} D^U \mid U \subseteq N_1 \}$
	    of functions between domains,
	    which satisfy
	    $\phi_V \cdot \pi_{V,U} = \pi_{fV,fU} \cdot \phi_U$
	    for every pair $U \subseteq V$ of subsets of $N_1$.
\end{enumerate}
Given any sorted domain $\pair{N}{D}$,
the full $N$th power $D^N$,
the unique constant function $N \stackrel{{\bf !}}{\rightarrow} {\bf 1}$, and
the collection of all projection functions
$\pi = \{ D^N \stackrel{\pi_{N,U}}{\rightarrow} D^U \mid U \subseteq N \}$
form a canonical morphism of sorted domains called {\em projection\/}
\[ \pi_{N,D}
   \define
   \pair{{\bf !}}{\pi} \type \pair{N}{D} \rightarrow \pair{{\bf 1}}{D^N}. \]

Morphisms of sorted domains
specify adjoint (direct/inverse image) relational flow.
Let
$\pair{f}{\phi} \type \pair{N_1}{D_1} \rightarrow \pair{N_2}{D_2}$
be any morphism of sorted domains. Define direct and inverse image operators as follows.
\begin{description}
	\item[{[Direct image]}] Let $R_2 \type \pair{E_2}{\tau_2}$ be any distributed
		$\pair{N_2}{D_2}$-relation.
		Define scheme $\pair{E_1}{\tau_1}$ to be the pullback
		of scheme $\pair{E_1}{\tau_1}$ along the direct image sort map
		$\wp f \type \wp N_1 \rightarrow \wp N_2$
		with fibered product
		$E_1 \define \{ (U,e_2) \mid U \subseteq N_1, e_2 \memberof E_2, fU = \tau_{2,e_2} \}$
		and projection function
		$\tau_1 \type (U,e_2) \mapsto U$.
		Define distributed relation
		$R_{1,(U,e_2)} \define \phi_U(R_{2,e_2})$
		for any $(U,e_2) \in E_1$
		to be the direct image along domain function
		$D_2^{\tau_{2,e_2}} = D^{fU} \stackrel{\phi_U}{\longrightarrow} D_1^U$.
		Finally,
		define the direct image distributed $\pair{N_1}{D_1}$-relation
		${\pair{f}{\phi}}_\ast(R_2 \type \pair{E_2}{\tau_2})
		  \define R_1 \type \pair{E_1}{\tau_1}$.
		Direct image is a monotonic function
		\[ {\pair{f}{\phi}}_\ast \type {\bf Rel}_{N_2,D_2} \rightarrow {\bf Rel}_{N_1,D_1}. \]
	\item[{[Inverse image]}] Let $R_1 \type \pair{E_1}{\tau_1}$ be any distributed
		$\pair{N_1}{D_1}$-relation.
		Define distributed relation
		$R_{2,e_1} \define \phi_{\tau_{1,e_1}}^{-1}(R_{1,e_1})$
		for any $e_1 \in E_1$
		to be the inverse image along domain function
		$D_2^{f\tau_{1,e_1}} \stackrel{\phi_{\tau_{1,e_1}}}{\longrightarrow} D_1^{\tau_{1,e_1}}$.
		Finally,
		define the inverse image distributed $\pair{N_2}{D_2}$-relation
		${\pair{f}{\phi}}^\ast(R_1 \type \pair{E_1}{\tau_1}) \define R_2 \type \pair{E_2}{\tau_2}$.
		Inverse image is a monotonic function
		\[ {\pair{f}{\phi}}^\ast \type {\bf Rel}_{N_1,D_1} \rightarrow {\bf Rel}_{N_2,D_2}. \]
\end{description}

\begin{Proposition}
	Relational flow is adjoint:
	for any morphism of sorted domains
	$\pair{f}{\phi} \type \pair{N_1}{D_1} \rightarrow \pair{N_2}{D_2}$,
	direct and inverse image form an adjoint pair of monotonic functions
	\[ {\pair{f}{\phi}}_\ast \;\dashv\; {\pair{f}{\phi}}^\ast \]
\end{Proposition}

\begin{Proposition}
	The satisfaction context of a distributed relation
	is the inverse image along projection
	\[ {\cal C}_{\cal R} = \pi_{N,D}^\ast(R \type \pair{E}{\tau}). \]
\end{Proposition}
The satisfaction context of a context
(regarded as a single-sorted distributed relation)
is the original context.
This demonstrates that distributed relations are a proper extension of formal contexts.



\appendix

\section{Networks of Constraints}

The basic constituents in constraint-based reasoning
are
tuples of values representing real-world objects,
and ways of describing these in terms of constraining relations.
The relationship between tuples and constraints is a derived relationship called {\em satisfaction\/}.
Constraint-based reasoning
is based upon the primitive notion of a network of constraints.
A {\em network of constraints\/} \cite{MR88} is a quintuple
${\cal N} = \quintuple{N}{\tau}{E}{D}{R}$
consisting of the following.
\begin{enumerate}
	\item A specification or syntactic part,
	 	represented by a hypergraph $\Omega = \triple{N}{\tau}{E}$,
	 	with
		a set $N$ of nodes called domain indices, sorts or variables; and
		a set $E$ of hyper-edges called constraint indices or just constraints,
		such that each edge $e \memberof E$ is typed $e \type I$ by a subset of domain indices
		$\tau_e \define I \subseteq N$ called its connection or scheme.
		The scheme is a subset function $\tau \type E \rightarrow {\wp}N$
		where ${\wp}N$ is the power-set of $N$;
		or equivalently,
		a relation $\tau \subset \product{N}{E}$ with $i{\tau}e$
		when $i$ indexes a domain of the constraint $e$.
		(So the triple $\triple{N}{E}{\tau}$ is a formal context as defined below,
		representing the ``syntactic concepts'' of the constraint network ${\cal N}$.
		These formal syntactic concepts consist of
		the intersections of schemes as intent and projectable constraints as extent.)
		In database theory elements of $N$ would be called attributes;
		but this terminology would be confusing here 
		since formal concept analysis has attributes
		which correspond here to elements of $E$.
		Sorts are identified with variables here.
		In order to separate these,
		the hypergraph $\Omega = \triple{N}{\tau}{E}$
		must be generalized to a multisorted first order signature.
	\item An interpretative or semantic part,
		which interprets the hypergraph,
		as
		an $N$-indexed collection $\{ D_i \mid i \memberof N \}$ of sets of values called domains,
		or an $N$-sorted domain;
		and
		an $E$-indexed collection
		$\{ R_e \subseteq D^I \mid e \memberof E, I = \tau_e \}$
		of constraint relations,
		where $D^I \define \prod_{i \in I} D_i$ is
		the Cartesian product of all domains with indices in $I$.
\end{enumerate}

\section{Example}

Table~\ref{network} gives an interesting network of constraints on Boolean values
as originally described in \cite{MR88}.
The variables (domain indices) $N = \{ a_1, a_2, a_3, a_4, a_5 \}$
represent the Boolean-valued domain
$D_i = \{ {\tt f}, {\tt t} \}$ for $1 \leq i \leq 5$.
The constraints $E = \{ e_1, e_2, e_3, e_4, e_5 \}$
are presented with schema and relational extents.
\scriptsize
\begin{table}
\begin{center}
	\begin{tabular}{cc}
	$\begin{array}{|r@{\;=\;}l|} \hline
		\multicolumn{2}{|c|}{\bf schema} \\ \hline\hline
		\tau_{e_1} & \{ a_1 \} \\
		\tau_{e_2} & \{ a_1, a_2, a_3 \} \\
		\tau_{e_3} & \{ a_1, a_4 \} \\
		\tau_{e_4} & \{ a_2, a_5 \} \\
		\tau_{e_5} & \{ a_3, a_4, a_5 \} \\ \hline
	 \end{array}$
	&
	$\begin{array}{|r@{\;=\;}l|} \hline
		\multicolumn{2}{|c|}{\bf constraint\mbox{  }relations} \\ \hline\hline
		R_{e_1} & \{ ({\rm f}) \} \\
		R_{e_2} & \{ ({\rm f},{\rm f},{\rm t}), ({\rm f},{\rm t},{\rm t}), ({\rm t},{\rm f},{\rm f}), ({\rm t},{\rm t},{\rm t}) \} \\
		R_{e_3} & \{ ({\rm f},{\rm t}), ({\rm t},{\rm t}), ({\rm f},{\rm f}) \} \\
		R_{e_4} & \{ ({\rm f},{\rm f}), ({\rm t},{\rm f}), ({\rm t},{\rm t}) \} \\
		R_{e_5} & \{ ({\rm f},{\rm f},{\rm f}), ({\rm f},{\rm t},{\rm t}), ({\rm t},{\rm t},{\rm f}) \} \\ \hline
	 \end{array}$
	\end{tabular}
\end{center}
\caption{{\bf Network of Constraints} ${\cal N}$ \label{network}}
\end{table}
\normalsize

Table~\ref{context} represents the formal satisfaction context
of the network of constraints presented in Table~\ref{network}
--- the fact $x{\models}e$ that the tuple $x$ satisfies the constraint $e$
is indicated by a `$\times$' in the ${x}{e}$-th entry.
\scriptsize
\begin{table}
\begin{center}
	\begin{tabular}{|l|c@{\hspace{3pt}}c@{\hspace{3pt}}c@{\hspace{3pt}}c@{\hspace{3pt}}c|} \hline
		            & $e_1$ & $e_2$ & $e_3$ & $e_4$ & $e_5$ \\ \hline
		(f,f,f,f,f) &$\times$&&$\times$&$\times$&$\times$ \\
		(f,f,f,f,t) &$\times$&&$\times$&& \\
		(f,f,f,t,f) &$\times$&&$\times$&$\times$& \\
		(f,f,f,t,t) &$\times$&&$\times$&&$\times$ \\
		(f,f,t,f,f) &$\times$&$\times$&$\times$&$\times$& \\
		(f,f,t,f,t) &$\times$&$\times$&$\times$&& \\
		(f,f,t,t,f) &$\times$&$\times$&$\times$&$\times$&$\times$ \\
		(f,f,t,t,t) &$\times$&$\times$&$\times$&& \\
		(f,t,f,f,f) &$\times$&&$\times$&$\times$& \\
		(f,t,f,f,t) &$\times$&&$\times$&$\times$& \\
		(f,t,f,t,f) &$\times$&&$\times$&$\times$& \\
		(f,t,f,t,t) &$\times$&&$\times$&$\times$&$\times$ \\
		(f,t,t,f,f) &$\times$&$\times$&$\times$&$\times$& \\
		(f,t,t,f,t) &$\times$&$\times$&$\times$&$\times$& \\
		(f,t,t,t,f) &$\times$&$\times$&$\times$&$\times$&$\times$ \\
		(f,t,t,t,t) &$\times$&$\times$&$\times$&$\times$& \\
		(t,f,f,f,f) &&$\times$&&$\times$&$\times$ \\
		(t,f,f,f,t) &&$\times$&&& \\
		(t,f,f,t,f) &&$\times$&$\times$&$\times$& \\
		(t,f,f,t,t) &&$\times$&$\times$&&$\times$ \\
		(t,f,t,f,f) &&&&$\times$& \\
		(t,f,t,f,t) &&&&& \\
		(t,f,t,t,f) &&&$\times$&$\times$&$\times$ \\
		(t,f,t,t,t) &&&$\times$&& \\
		(t,t,f,f,f) &&&&$\times$&$\times$ \\
		(t,t,f,f,t) &&&&$\times$& \\
		(t,t,f,t,f) &&&$\times$&$\times$& \\
		(t,t,f,t,t) &&&$\times$&$\times$&$\times$ \\
		(t,t,t,f,f) &&$\times$&&$\times$& \\
		(t,t,t,f,t) &&$\times$&&$\times$& \\
		(t,t,t,t,f) &&$\times$&$\times$&$\times$&$\times$ \\
		(t,t,t,t,t) &&$\times$&$\times$&$\times$& \\ \hline
	\end{tabular}
\end{center}
\caption{{\bf Satisfaction Context} ${\cal C}_{\cal N}$ \label{context}}
\end{table}
\normalsize

There are two possible projective containment conditions
$\{ e_2 \leq e_1, e_3 \leq e_1 \}$
for this network.
Although
neither of these is satisfied by the network presented in Table~\ref{network}
both are satisfied by the network's interior.
The solution set is the pair of quintuples
\footnotesize
\begin{center}
	$\begin{array}{|r@{\;=\;}l|} \hline
		\multicolumn{2}{|c|}{\bf solution\mbox{  }set} \\ \hline\hline
		{\Pi}R & \{ ({\rm f},{\rm f},{\rm t},{\rm t},{\rm f}),
		            ({\rm f},{\rm t},{\rm t},{\rm t},{\rm f})
		             \} \\ \hline
	 \end{array}$
\end{center}
\normalsize

Table~\ref{interior} gives
the interior within the network of constraints presented in Table~\ref{network}.
\scriptsize
\begin{table}
\begin{center}
	\begin{tabular}{cc}
	$\begin{array}{|r@{\;=\;}l|} \hline
		\multicolumn{2}{|c|}{\bf schema} \\ \hline\hline
		\tau_{e_1} & \{ a_1 \} \\
		\tau_{e_2} & \{ a_1, a_2, a_3 \} \\
		\tau_{e_3} & \{ a_1, a_4 \} \\
		\tau_{e_4} & \{ a_2, a_5 \} \\
		\tau_{e_5} & \{ a_3, a_4, a_5 \} \\ \hline
	 \end{array}$
	&
	$\begin{array}{|r@{\;=\;}l|} \hline
		\multicolumn{2}{|c|}{\bf constraint\mbox{  }relations} \\ \hline\hline
		{{\pi}{\Pi}R}_{e_1} & \{ ({\rm f}) \} \\
		{{\pi}{\Pi}R}_{e_2} & \{ ({\rm f},{\rm f},{\rm t}), ({\rm f},{\rm t},{\rm t}) \} \\
		{{\pi}{\Pi}R}_{e_3} & \{ ({\rm f},{\rm t}) \} \\
		{{\pi}{\Pi}R}_{e_4} & \{ ({\rm f},{\rm f}), ({\rm t},{\rm f}) \} \\
		{{\pi}{\Pi}R}_{e_5} & \{ ({\rm t},{\rm t},{\rm f}) \} \\ \hline
	 \end{array}$
	\end{tabular}
\end{center}
\caption{{\bf Interior System} ${\cal N}^{{\circ}}$ \label{interior}}
\end{table}
\normalsize
This was erroneously described in \cite{MR88}.
The error may have caused Montanari and Rossi to overlook
the important projective containment conditions for systems of constraints,
which lead to the definition of distributed relations.
In this example we see,
as \cite{MR88} have pointed out,
why distributed relations are more optimal w.r.t.\ storage requirements than networks:
although ${\pi}{\Pi}R$ is equivalent to $R$,
it has only half as many tuples.
In addition,
there are only half as many concepts in the concept lattice of the interior.

This paper takes the position that
the hidden semantic structure of the system of constraints
inside any distributed relation ${\cal R}$
(and in particular, any network of constraints ${\cal N}$)
is revealed by
the concept lattice ${\bf L}({\cal C}_{\cal R})$ of the formal satisfaction context
associated with the system.
For any collection of constraint indices $\psi \subseteq E$,
the inverse derivation
\footnotesize
\begin{center}
	$\begin{array}{r@{\;\;=\;\;}l}
		\deriveinv{\psi}{\sigma}
		& \{ x \memberof D^N \mid x{\sigma}e \mbox{ for all constraints } e \in \psi \} \\
		& \{ x \memberof D^N \mid \pi_{\tau_e}(x) \memberof R_e \mbox{ for all constraints } e \in \psi \}
	 \end{array}$
\end{center}
\normalsize
corresponds to the natural database join ${\Join}\{ R_e \mid e \memberof \psi \}$
of the constraint relations indexed in the set $\psi$.
The most efficient way to describe concepts in any concept lattice
is by means of generators and successors:
the extent of a concept is the set of tuple generators
of all predeceeding concepts (concepts below it);
and
the intent of a concept is the set of constraint generators
of all succeeding concepts (concepts above it).
The 23 concepts
for the satisfaction context of the network of constraints
presented in Table~\ref{network},
are described in Table~\ref{generators} and Table~\ref{successors}
by listing their generators and successors.
Concepts are indexed in lectic (reverse numeric) order
of their satisfaction bit vector for constraint indices.
The top of the lattice is $C_1 = \pair{D^N}{\emptyset}$
and is label by the single tuple $(t,f,t,f,t)$ which satisfies no constraint.
The bottom of the lattice is $C_{23} = \pair{{\Pi}R}{E}$
and is label by the solution set $\{ (f,f,t,t,f), (f,t,t,t,f) \}$,
the two tuples which satisfy all constraints.
From Table~\ref{generators} and Table~\ref{successors}
we can compute extents and intents of concepts:
for example,
the concept $C_{10}$ has extent
\footnotesize 
$\{ (t,f,f,f,f), (t,f,f,t,t), (f,f,t,t,f), (f,t,t,t,f), (t,t,t,t,f) \}$
\normalsize
and intent
$\{ e_2, e_5 \}$.
Here, as for all concepts,
the extent is the solution set for the constraints in the intent,
and the intent is {\em all\/} constraints satisfied by this solution set
--- they determine each other.
The extent is also the natural database join $R_{e_2} {\Join} R_{e_5}$
of the constraint relations indexed in the intent.
\scriptsize
\begin{table}
\begin{center}
	\begin{tabular}{|c|l|c|} \hline
		& \multicolumn{2}{|c|}{\bf generators} \\ \cline{2-3}
		\raisebox{4pt}{\bf concept} & \multicolumn{1}{|c|}{\bf tuples} & {\bf constraints} \\ \hline
		$C_1$     & (t,f,t,f,t) & \\
		$C_2$     &             & $e_5$ \\
		$C_3$     & (t,f,t,f,f), (t,t,f,f,t) & $e_4$ \\
		$C_4$     & (t,t,f,f,f) &       \\
		$C_5$     & (t,f,t,t,t) & $e_3$ \\
		$C_6$     &  & \\
		$C_7$     & (t,t,f,t,f) & \\
		$C_8$     & (t,f,t,t,f), (t,t,f,t,t) & \\
		$C_9$     & (t,f,f,f,t) & $e_2$ \\
		$C_{10}$  &  & \\
		$C_{11}$  & (t,t,t,f,f), (t,t,t,f,t) & \\
		$C_{12}$  & (t,f,f,f,f) & \\
		$C_{13}$  &  & \\
		$C_{14}$  & (t,f,f,t,t) & \\
		$C_{15}$  & (t,f,f,t,f), (t,t,t,t,t) & \\
		$C_{16}$  & (t,t,t,t,f) & \\
		$C_{17}$  & (f,f,f,f,t) & $e_1$\\
		$C_{18}$  & (f,f,f,t,t) & \\
		$C_{19}$  & (f,f,f,t,f), (f,t,f,f,f), (f,t,f,f,t), (f,t,f,t,f) & \\
		$C_{20}$  & (f,f,f,f,t), (f,t,f,t,t) & \\
		$C_{21}$  & (f,t,t,f,t), (f,f,t,t,t) & \\
		$C_{22}$  & (f,f,t,f,f), (f,t,t,f,f), (f,t,t,f,t), (f,t,t,t,t) & \\
		$C_{23}$  & (f,f,t,t,f), (f,t,t,t,f) & \\ \hline
	\end{tabular}
\end{center}
\caption{{\bf Concept Generators} \label{generators}}
\end{table}
\normalsize

\scriptsize
\begin{table}
\begin{center}
	\begin{tabular}{|c|l|} \hline
		{\bf concept} & \multicolumn{1}{c|}{{\bf successor concepts}} \\ \hline\hline
		$C_1$    & $\emptyset$ \\
		$C_2$    & $\{ C_{1} \}$ \\
		$C_3$    & $\{ C_{1} \}$ \\
		$C_4$    & $\{ C_{2}, C_{3} \}$ \\
		$C_5$    & $\{ C_{1} \}$ \\
		$C_6$    & $\{ C_{2}, C_{5} \}$ \\
		$C_7$    & $\{ C_{3}, C_{5} \}$ \\
		$C_8$    & $\{ C_{4}, C_{6}, C_{7} \}$ \\
		$C_9$    & $\{ C_{1} \}$ \\
		$C_{10}$ & $\{ C_{2}, C_{9} \}$ \\
		$C_{11}$ & $\{ C_{3}, C_{9} \}$ \\
		$C_{12}$ & $\{ C_{4}, C_{10}, C_{11} \}$ \\
		$C_{13}$ & $\{ C_{5}, C_{9} \}$ \\
		$C_{14}$ & $\{ C_{6}, C_{10}, C_{13} \}$ \\
		$C_{15}$ & $\{ C_{7}, C_{11}, C_{13} \}$ \\
		$C_{16}$ & $\{ C_{8}, C_{12}, C_{14}, C_{15} \}$ \\
		$C_{17}$ & $\{ C_{5} \}$ \\
		$C_{18}$ & $\{ C_{6}, C_{17} \}$ \\
		$C_{19}$ & $\{ C_{7}, C_{17} \}$ \\
		$C_{20}$ & $\{ C_{8}, C_{18}, C_{19} \}$ \\
		$C_{21}$ & $\{ C_{13}, C_{17} \}$ \\
		$C_{22}$ & $\{ C_{15}, C_{19}, C_{21} \}$ \\
		$C_{23}$ & $\{ C_{16}, C_{20}, C_{22} \}$ \\ \hline
	\end{tabular}
\end{center}
\caption{{\bf Concept Successors} \label{successors}}
\end{table}
\normalsize
Table~\ref{order} represents the order relation for the concept lattice.
The matrix is lower triangular because of two facts:
(1) the concepts are listed in lectic (reverse numeric) order; and
(2) the concept lattice partial order is a suborder of the lectic total order.
\scriptsize
\begin{table}
\begin{center}
	\begin{tabular}{|l|c@{\hspace{2pt}}c@{\hspace{2pt}}c@{\hspace{2pt}}c@{\hspace{2pt}}c@{\hspace{2pt}}c@{\hspace{2pt}}c@{\hspace{2pt}}c@{\hspace{2pt}}c@{\hspace{2pt}}c@{\hspace{1pt}}c@{\hspace{1pt}}
	                   c@{\hspace{1pt}}c@{\hspace{1pt}}c@{\hspace{1pt}}c@{\hspace{1pt}}c@{\hspace{1pt}}c@{\hspace{1pt}}c@{\hspace{1pt}}c@{\hspace{1pt}}c@{\hspace{1pt}}c@{\hspace{1pt}}c@{\hspace{1pt}}c|} \hline
		         &$1$&$2$&$3$&$4$&$5$&$6$&$7$&$8$&$9$&${10}$&${11}$&${12}$&${13}$&${14}$&${15}$&${16}$&${17}$&${18}$&${19}$&${20}$&${21}$&${22}$&${23}$ \\ \hline
		$1$    &$\times$&&&&&&&&&&&&&&&&&&&&&& \\
		$2$    &$\times$&$\times$&&&&&&&&&&&&&&&&&&&&& \\
		$3$    &$\times$&&$\times$&&&&&&&&&&&&&&&&&&&& \\
		$4$    &$\times$&$\times$&$\times$&$\times$&&&&&&&&&&&&&&&&&&& \\
		$5$    &$\times$&&&&$\times$&&&&&&&&&&&&&&&&&& \\
		$6$    &$\times$&$\times$&&&$\times$&$\times$&&&&&&&&&&&&&&&&& \\
		$7$    &$\times$&&$\times$&&$\times$&&$\times$&&&&&&&&&&&&&&&& \\
		$8$    &$\times$&$\times$&$\times$&$\times$&$\times$&$\times$&$\times$&$\times$&&&&&&&&&&&&&&& \\
		$9$    &$\times$&&&&&&&&$\times$&&&&&&&&&&&&&& \\
		${10}$ &$\times$&$\times$&&&&&&&$\times$&$\times$&&&&&&&&&&&&& \\
		${11}$ &$\times$&&$\times$&&&&&&$\times$&&$\times$&&&&&&&&&&&& \\
		${12}$ &$\times$&$\times$&$\times$&$\times$&&&&&$\times$&$\times$&$\times$&$\times$&&&&&&&&&&& \\
		${13}$ &$\times$&&&&$\times$&&&&$\times$&&&&$\times$&&&&&&&&&& \\
		${14}$ &$\times$&$\times$&&&$\times$&$\times$&&&$\times$&$\times$&&&$\times$&$\times$&&&&&&&&& \\
		${15}$ &$\times$&&$\times$&&$\times$&&$\times$&&$\times$&&$\times$&&$\times$&&$\times$&&&&&&&& \\
		${16}$ &$\times$&$\times$&$\times$&$\times$&$\times$&$\times$&$\times$&$\times$&$\times$&$\times$&$\times$&$\times$&$\times$&$\times$&$\times$&$\times$&&&&&&& \\
		${17}$ &$\times$&&&&$\times$&&&&&&&&&&&&$\times$&&&&&& \\
		${18}$ &$\times$&$\times$&&&$\times$&$\times$&&&&&&&&&&&$\times$&$\times$&&&&& \\
		${19}$ &$\times$&&$\times$&&$\times$&&$\times$&&&&&&&&&&$\times$&&$\times$&&&& \\
		${20}$ &$\times$&$\times$&$\times$&$\times$&$\times$&$\times$&$\times$&$\times$&&&&&&&&&$\times$&$\times$&$\times$&$\times$&&& \\
		${21}$ &$\times$&&&&$\times$&&&&$\times$&&&&$\times$&&&&$\times$&&&&$\times$&& \\
		${22}$ &$\times$&&$\times$&&$\times$&&$\times$&&$\times$&&$\times$&&$\times$&&$\times$&&$\times$&&$\times$&&$\times$&$\times$& \\
		${23}$ &$\times$&$\times$&$\times$&$\times$&$\times$&$\times$&$\times$&$\times$&$\times$
		         &$\times$&$\times$&$\times$&$\times$&$\times$&$\times$&$\times$&$\times$&$\times$&$\times$&$\times$&$\times$&$\times$&$\times$ \\ \hline
	\end{tabular}
\end{center}
\caption{{\bf Order Relation for Concept Lattice} ${\bf L}({\cal C}_{\cal N})$ \label{order}}
\end{table}
\normalsize

\section{Relational Interior}

Any distributed relation contains within it
another distributed relation called its interior
which satisfies all possible projective containment conditions.
Interior is derived from
an interesting adjointness between projection and solution.

On the one hand,
the {\em solution set\/} of a distributed relation
$R \in {\rm Rel}_{N,D}^{E,\tau}$
is defined by
${\Pi}R \define \{ x \memberof D^N \mid x{\models}e \;\mbox{for all}\; e \memberof E \}$.
Two distributed relations $R,S \in {\rm Rel}_{N,D}^{E,\tau}$
are {\em equivalent\/}
when they have the same solution set ${\Pi}R = {\Pi}S$.
Define
${\Pi}^{-1}P \define \{ R \mid {\Pi}R = P \}$
to be the (possibly empty) collection of distributed relations with solution set $P$.
By definition of the solution set ${\Pi}R$,
the projection $\pi_{N{\tau_e}}(x)$ of every tuple $x \memberof {\Pi}R$
is contained in the $e$-th constraint relation $\pi_{N{\tau_e}}(x) \memberof R_e$.
This fact is expressed by the ``counit inequality''
${\pi}({\Pi}R) \subseteq R$.

On the other hand,
let $P \subseteq D^N$ be any subset of full object tuples.
We regard $P$ as a potential solution set.
The {\em projection\/} of $P$
is the $E$-indexed collection of relations
${\pi}P \define \{ \pi_{N{\tau_e}}(P) \mid e \memberof E \}$.
The projection of $P$
clearly satisfies the projective containment conditions
$\pi_{{\tau_e}{\tau_d}}(\pi_{N{\tau_e}}(P)) \subseteq \pi_{N{\tau_d}}(P)$,
and hence defines a distributed relation
$\quintuple{N}{\tau}{E_{\tau}}{D}{{\pi}P}$.
The fact that every tuple in $P$ satisfies every constraint in $E$
is expressed by the ``unit inequality''
$P \subseteq {\Pi}({\pi}P)$.

There are two ways to define a context on full tuples.
\begin{enumerate}
	\item For any subset of full tuples $P \subseteq D^N$,
		define an associated order-theoretic formal context
		${\cal C}_P = \triple{G}{M}{I_P}$
		by
		\footnotesize
		\[ x{I_P}e \mbox{  iff  } x \memberof P. \]
		\normalsize
\end{enumerate}

Since the projection $\pi$ and product $\Pi$ operators are monotonic functions,
we have the following proposition.
\begin{Proposition}
	The projection operator $\pi$
	and
	the solution operator $\Pi$
	form an adjointness or Galois connection
	\[ \pi \dashv \Pi \]
	between distributed relations $R$ and potential solution sets $P$,
	expressed by the logical equivalence:
	${\pi}P \subseteq R$ iff $P \subseteq {\Pi}R$.
\end{Proposition}
By the above adjointness,
projection preserves joins
${\pi}(\cup_k P_k) = \sqcup_k {\pi}(P_k)$,
and solution preserves meets
${\Pi}(\sqcap_k R_k) = \cap_k {\Pi}P_k$.

For any distributed relation
$R \in {\rm Rel}_{N,D}$
with distributed scheme $\pair{E}{\tau}$,
the projection of the solution
$R^{\circ} \define {\Pi}{\pi}(R) = {\pi}({\Pi}(R))$
is called the {\em interior\/} of $R$.
The interior is a distributed relation
which satisfies all possible projective containment conditions.
The above adjointness implies
the existence of minimal equivalent networks of constraints \cite{MR88}.
The interior $R^{\circ}$ is optimal w.r.t.\ satisfaction
--- it is the smallest distributed relation having solution set ${\Pi}R$:
\begin{enumerate}
	\item ${\Pi}(R^{\circ}) = {\Pi}R$; and
	\item if $S$ is an $E$-indexed distributed relations
		having solution set ${\Pi}S = {\Pi}R$,
		then $R^{\circ} \subseteq S$.
\end{enumerate}
In particular,
interior is the minimal distributed relation 
$R^{\circ} = \bigwedge {\Pi}^{-1} {\Pi}R$
equivalent to $R$.
These results were stated and proved in \cite{MR88},
but the proofs were not expressed
by use of the simplifying notion of an adjointness.

\section{Relational Participation}

The previous discussion
leads one to ask why certain tuples are not included in
the interior of a distributed relation $R$.
For any relational constraint index $e \memberof E$
a tuple $x \type \tau_e$
is in the difference
$R_e \setminus R^{\circ}_e$
when it does not participate
in a combining relationship with other tuples
to form a full object tuple that respects all constraints in $E$.
In other words,
it is isolated with respect to $R$.
In this section we will relativize this idea of {\em relational participation\/}
and make it more explicit.

We can emphasize the importance of certain concepts in a concept lattice
${\bf L}({\cal C})$
for a formal context ${\cal C} = \triple{G}{M}{I}$
by specifying this collection of concepts as a suborder
$P \subseteq {\bf L}({\cal C})$.
More precisely,
suppose that
$P \stackrel{\iota}{\rightarrow} {\bf L}({\cal C})$
is an injective monotonic function
which is left adjoint to
${\bf L}({\cal C}) \stackrel{\iota^\dashv}{\rightarrow} P$.
The trivial example of such a suborder is the full lattice,
where $\iota$ and $\iota^\dashv$ are both identity.
The simplest nontrivial example of such a suborder is the principal ideal
${\downarrow}c$ of any concept $c \memberof {\bf L}({\cal C})$,
where $\iota$ is subset inclusion $\iota(c') = c'$ for all $c' \leq_{\rm L} c$
and $\iota^\dashv$ is meet $\iota^\dashv(d) \define c \wedge_{\rm L} d$ for all $d \memberof {\bf L}({\cal C})$.
As a special case, relevant for relational interior, is $c = \bot$.
Define an associated formal context
${\cal C}_P = \triple{G}{M}{I_P}$
by
\footnotesize
\[ g{I_P}m \mbox{  iff  } \exists_{p \in P}\; \gamma(g) \leq_{\rm L} \iota(p) \leq_{\rm L} \mu(m). \]
\normalsize

\end{document}